# A Computational Economy for Grid Computing and its Implementation in the Nimrod-G Resource Broker

David Abramson §, Rajkumar Buyya §, and Jonathan Giddy †

§ School of Computer Science and Software Engineering
Monash University, Caulfield Campus
Melbourne, Australia

† CRC for Distributed Systems Technology
Monash University, Caulfield Campus
Melbourne, Australia

Email: {davida, rajkumar, jon}@csse.monash.edu.au

**Abstract:** Computational Grids, coupling geographically distributed resources such as PCs, workstations, clusters, and scientific instruments, have emerged as a next generation computing platform for solving large-scale problems in science, engineering, and commerce. However, application development, resource management, and scheduling in these environments continue to be a complex undertaking. In this article, we discuss our efforts in developing a resource management system for scheduling computations on resources distributed across the world with varying quality of service. Our service-oriented grid computing system called Nimrod-G manages all operations associated with remote execution including resource discovery, trading, scheduling based on economic principles and a user defined quality of service requirement. The Nimrod-G resource broker is implemented by leveraging existing technologies such as Globus, and provides new services that are essential for constructing industrial-strength Grids. We discuss results of preliminary experiments on scheduling some parametric computations using the Nimrod-G resource broker on a world-wide grid testbed that spans five continents.

## 1. Introduction

The accelerated development of Grid computing systems has positioned them as promising next generation computing platforms. They enable the coordinated use of geographically distributed resources, often owned by autonomous organizations, for creating virtual enterprises for solving large-scale problems in science, engineering, and commerce [1][7]. However, application composition, resource management and scheduling in these environments is a complex undertaking. This is due to the geographic distribution of resources that are often owned by different organizations having different usage policies and cost models, and varying loads and availability patterns. To address these resource management challenges, we have proposed and developed a computational economy framework for resource allocation and regulation of supply and demand for resources. The new framework offers incentive to resource owners for being part of the Grid and motivates resource users to trade off between time for results delivery and economic cost, i.e., deadline and budget [5].

We are exploring the use of an economic paradigm for Grid computing. We have developed an economy driven grid resource broker within the Nimrod-G system that supports soft-deadline and budget based scheduling of applications on the computational Grid [7]. Depending on users' Quality of Service (QoS) requirements, our resource broker dynamically leases Grid services at runtime depending on their cost, quality, and availability. The scheduler allows minimisation of time or cost within specific deadline and budget constraints.

Resource management systems need to provide mechanisms and tools that realize the goals of both service providers and consumers. The resource consumers need a *utility model*, representing their resource demand and preferences, and *brokers* that automatically generate strategies for choosing providers based on this model. Further, the brokers need to manage all issues associated with the execution of the underlying application. The service providers need *price generation schemes* so as to increase system utilization, as well as economic *protocols* that help them to offer competitive services. For the market to be competitive and efficient, coordination mechanisms are required that help the market reach an equilibrium price, that is, the market price at which the supply of a service equals the quantity demanded [13]. Numerous economic theories have been proposed in the literature and many commonly used economic models for selling goods and services can be employed as negotiation protocols in Grid computing. Some of these market or social driven economic models are shown in Table 1 along with the identity of the distributed system that adopted the approach [8].

| Economic Model | Adopted by |
| --- | --- |
| Commodity Market | Mungi [16], MOSIX [17], & Nimrod-G [1][4] |
| Posted Price | Nimrod-G |
| Bargaining | Mariposa [12] & Nimrod-G |
| Tendering or Contract-Net Model | Mariposa [12] |
| Auction Model | Spawn [18] & Popcorn [19] |
| Bid-based Proportional Resource Sharing | Rexec & Anemone [20] |
| Community, Coalition, and Bartering | Condor, SETI@Home [21], & MojoNation [22] |
| Monopoly and Oligopoly | Nimrod-G broker can be used to choose between resources offered at different quality and prices. |

**Table 1:** Economics models and example distributed computing scheduling systems that adopted them.

These economic models regulate the supply and demand for resources in Grid-based virtual enterprises. We demonstrate the power of these models in scheduling computations using the Nimrod-G resource broker on a grid testbed, called the World Wide Grid (WWG) spanning across five continents. Whilst it is not the goal of the system to earn revenue for the resource providers, this approach does provide an economic incentive for resource owners to share their resources on the Grid. Further, it encourages the emergence of a new service oriented computing industry. Importantly, it provides mechanisms to trade-off QoS parameters, deadline and computational cost, and offers incentive for users to relax their requirements. For example, a user may be prepared to accept a later deadline if the computation can be achieved more cheaply.

## 2. Nimrod-G: Economics driven Grid Computing Environment

### *2.1. Objectives and goals*

Nimrod-G is a tool for automated modeling and execution of parameter sweep applications (parameter studies) over global computational grids [1][2][3]. It provides a simple declarative parametric modeling language for expressing parametric experiments. A domain expert can easily create a plan for a parametric experiment and use the Nimrod system to submit jobs for execution. It uses novel resource management and scheduling algorithms based on economic principles. Specifically, it supports user-defined deadline and budget constraints for schedule optimisations and manages supply and demand of resources in the Grid using a set of resource trading services called GRACE (Grid Architecture for Computational Economy) [5][6][7].

Nimrod-G provides a persistent and programmable *task-farming engine* (TFE) that enables "plugging" *of* user-defined schedulers and customised applications or problem solving environments (e.g., ActiveSheets [24]) in place of default components. The task-farming engine is a coordination point for processes performing resource trading, scheduling, data and executable staging, remote execution, and result collation. In the past, the major focus of our project was on creating *tools* that help domain experts to compose their legacy serial applications for parameter studies and run them on computational clusters and manually managed grids [2][3]. Our current focus is on the use of economic principles in resource management and scheduling on the grid in order to provide some measurable quality of service to the end user. Real-world economic methods provide incentives for owners to contribute their resources to markets, and it also provides consumers with a basis for trading the quality of service they receive against cost. That is, our focus revolves within an intersection area of grid architectures, economic principles, and scheduling optimizations (see Figure 1), which is essential for pushing the grid into the mainstream computing.

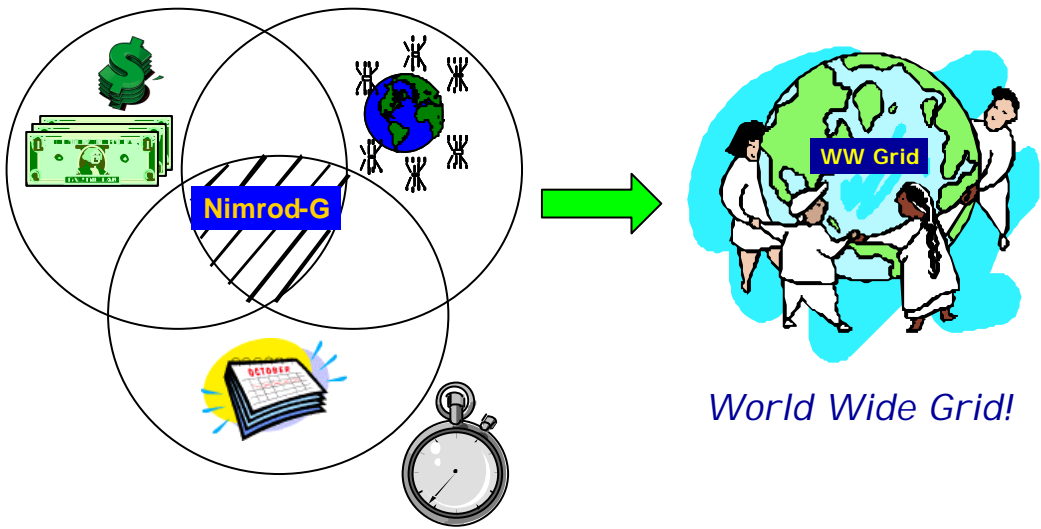

**Figure 1:** QoS based resource management—intersection of economic, scheduling, and grid worlds.

## 2.2. Services and End Users

Nimrod-G provides a suite of tools and services for creating parameter sweep applications, performing resource management, and scheduling applications. They are, a simple declarative programming language and associated GUI tools for creating scripts and parameterization of application input data files, and a grid resource broker with programmable entities for processing jobs on grid resources. The resource broker is made of a number of components, namely a persistent and programmable task farming engine, a schedule advisor, and a dispatcher, whose functionalities are discussed later. It also provides job management services that can be used for creating user-defined schedulers, steering and monitoring tools, and customized applications. Therefore, the end users that benefit from Nimrod-G tools, protocols, and services are:

- **Domain Experts:** This group includes scientific, engineering, and commercial users with large scale data-set processing requirements. Parameter applications can use Nimrod-G tools to compose them as coarse-grained data-parallel, parameter sweep applications for executing on distributed resources. They can also take advantage of the Nimrod-G broker features to trade off between a deadline and the cost of computation while scheduling application execution on the grid. This quality of service aspect is important to end users, because the results are only useful if they are returned in a timely manner. Previous Grid work has largely ignored this aspect of running real applications.

- **Problem Solving Environments Developers:** Application developers can grid enable their applications with their own mechanisms to submit jobs to the Nimrod-G resource broker at runtime depending on user requirements for processing on the Grid. This gives them the ability to create applications capable of directly using Nimrod-G tools and job management services, which in turn enables their applications for Grid execution.

- **Task Farming or Master-Worker Programming Environments Designers:** These users can focus on designing and developing easy use and powerful application creation primitives for task farming and master-work style programming model; developing translators and application execution environments by taking advantage of Nimrod-G runtime machinery for executing jobs on distributed grid resources.

- **Scheduling Researchers:** The scheduling policy developers generally use simulation techniques and tools such as GridSim [27] and Simgrid [28] for evaluating performance of their algorithms. In simulation, it is very difficult to capture the complete property and behavior of a real world system and hence, evaluation results may be inaccurate. Accordingly, to prove the usefulness of scheduling algorithms on actual systems, researchers need to develop runtime machinery, which is a resource intensive and time-consuming task. This can be overcome by using Nimrod-G broker programmable capability. Researchers can use Nimrod-G job management protocols and services to develop their own scheduler and associated scheduling algorithms. The new scheduler can be used to run actual applications on distributed resources and then evaluate ability of scheduling algorithms in optimally mapping jobs to resources.

## 2.3. Architecture

The Nimrod-G toolkit and resource broker is developed by leveraging services provided by Grid middleware systems such as Globus, Legion, Condor/G, and the GRACE trading mechanisms. These middleware systems provide a set of low-level protocols for secure and uniform access to remote resources; and services for accessing resources information and storage management,. The modular and layered architecture of Nimrod-G is shown in Figure 2. The key components of Nimrod-G resource broker consist of:

- Nimrod-G Clients, which can be:
    - Tools for creating parameter sweep applications.
    - Steering and control monitors, and
    - Customised end user applications (e.g., ActiveSheets [24]).
- The Nimrod-G Resource Broker, that consists of:
    - A Task Farming Engine (TFE),
    - A Scheduler that performs resource discovery, trading, and scheduling,
    - A Dispatcher and Actuator, and
    - Agents for managing the execution of jobs on resources.

The Nimrod-G broker architecture leverages services provided by lower-level different grid middleware solutions to perform resource discovery, trading, and deployment of jobs on grid resources.

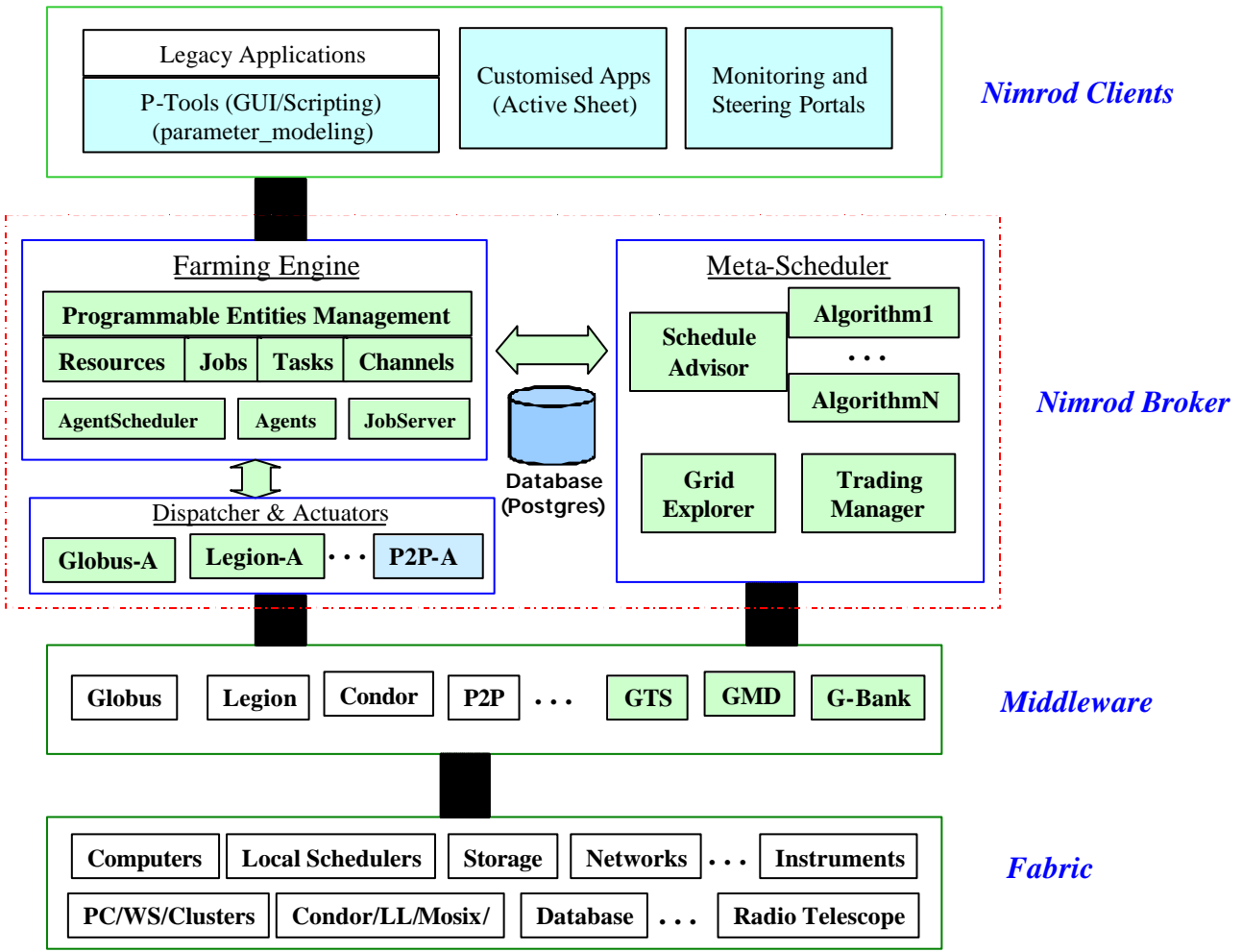

**Figure 2:** Layered with an hourglass shape architecture of Nimrod-G system.

### *2.3.1. Nimrod-G Clients*

*Tools for Creating Parameter Sweep Applications*

Nimrod supports GUI tools and declarative programming language that assist in creation of parameter sweep applications [2]. They allow user to: a) parameterise input files, b) prepare a plan file containing the commands that define parameters and their values c) generate a run file, which converts the generic plan file to a detailed list of jobs; and d) control and monitor execution of the jobs. The application execution environment handles online creation of input files and command line arguments through parameter substitution.

*Steering and Control Monitors*

These components act as a user-interface for controlling and monitoring a Nimrod-G experiment. The user can vary constraints related to time and cost that influence the direction the scheduler takes while selecting resources. It serves as a monitoring console and lists the status of all jobs, which a user can view and control. A Nimrod-G monitoring and steering client snapshot is shown in Figure 4. Another feature of the Nimrod-G client is that it is possible to run multiple instances of the same client at different locations. That means the experiment can be started on one machine, monitored on another machine by the same or different user, and the experiment can be controlled from yet another location. We have used this feature to monitor and control an experiment from Monash University and Pittsburgh Supercomputing Centre at Carnegie Melon University simultaneously during HPDC-2000 research demonstrations.

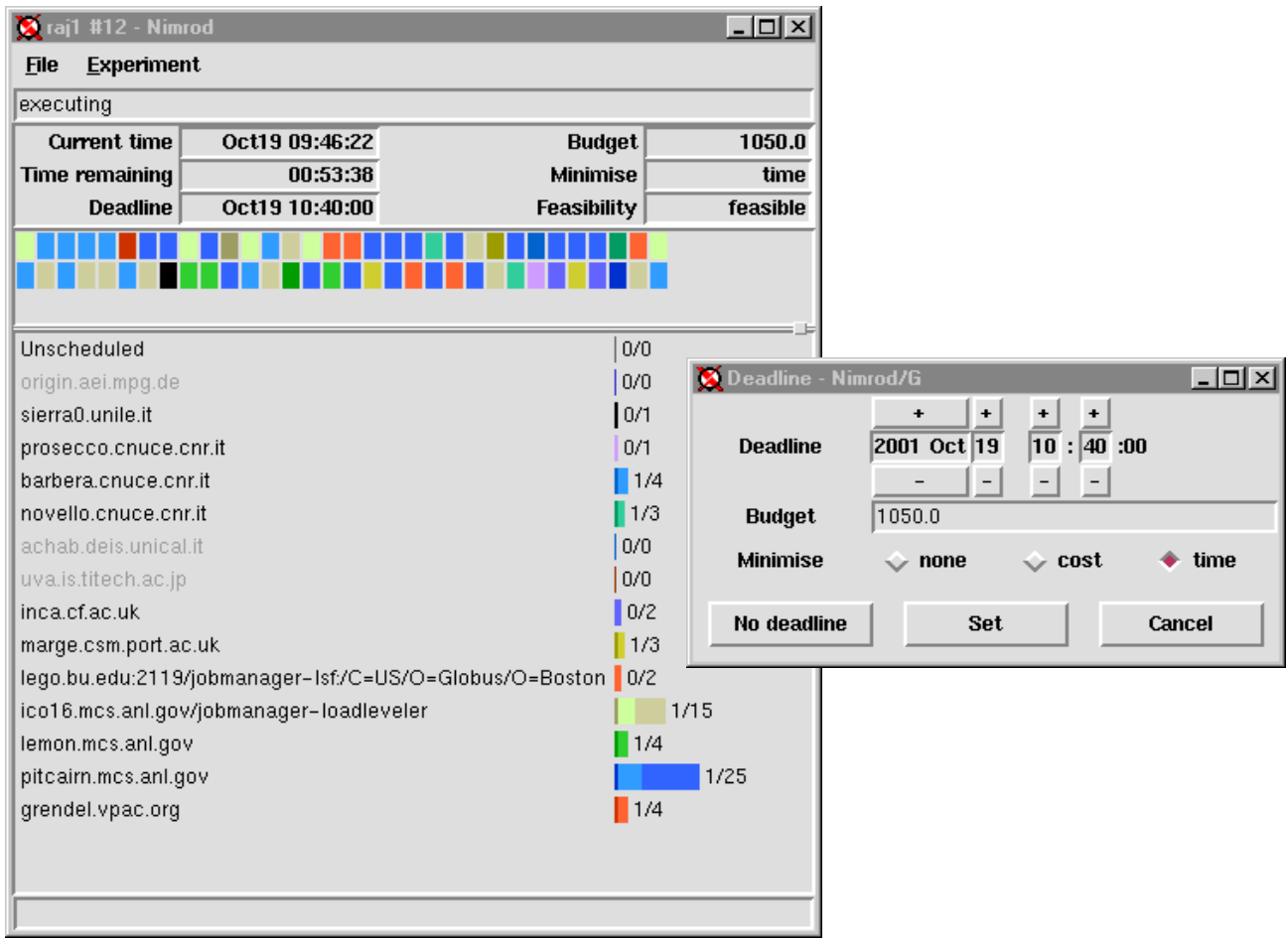

**Figure 3:** A Snapshot of Nimrod-G Execution Monitoring and Steering Client.

*Customised End User Applications*

Specialized applications can be developed to create jobs at runtime and add jobs to the Nimrod-G Engine for processing on the Grid. These applications can use the Nimrod-G management APIs for adding and managing jobs. One such application is ActiveSheets [24], an extended Microsoft Excel spreadsheet that submits cell functions for parallel execution on computational grids using Nimrod-G services. Another example is the Nimrod/O system, a tool that uses non-linear optimization algorithms to facilitate automatic optimal design [29]. This tool has been used on a variety of case studies, including antenna design, smog modeling, durability optimization, aerofoil design, computational fluid dynamics [30].

### 2.3.2. The Nimrod-G Grid Resource Broker

The Nimrod-G Resource broker is responsible for determining the specific requirements that an experiment places on the Grid and performing resource discovery, scheduling, dispatching jobs to remote Grid nodes, starting and managing job execution, and gathering results back to the home node. The sub-modules of our resource broker are, the task farming engine; the scheduler that consists of a grid explorer for resource discovery, a schedule advisor backed with scheduling algorithms, and a resource trading manager; a dispatcher and actuators for deploying agents on grid resources; and agents for managing execution of Nimrod-G jobs on grid resources. The interaction between components of the Nimrod-G runtime machinery and Grid services during runtime is shown in Figure 4.

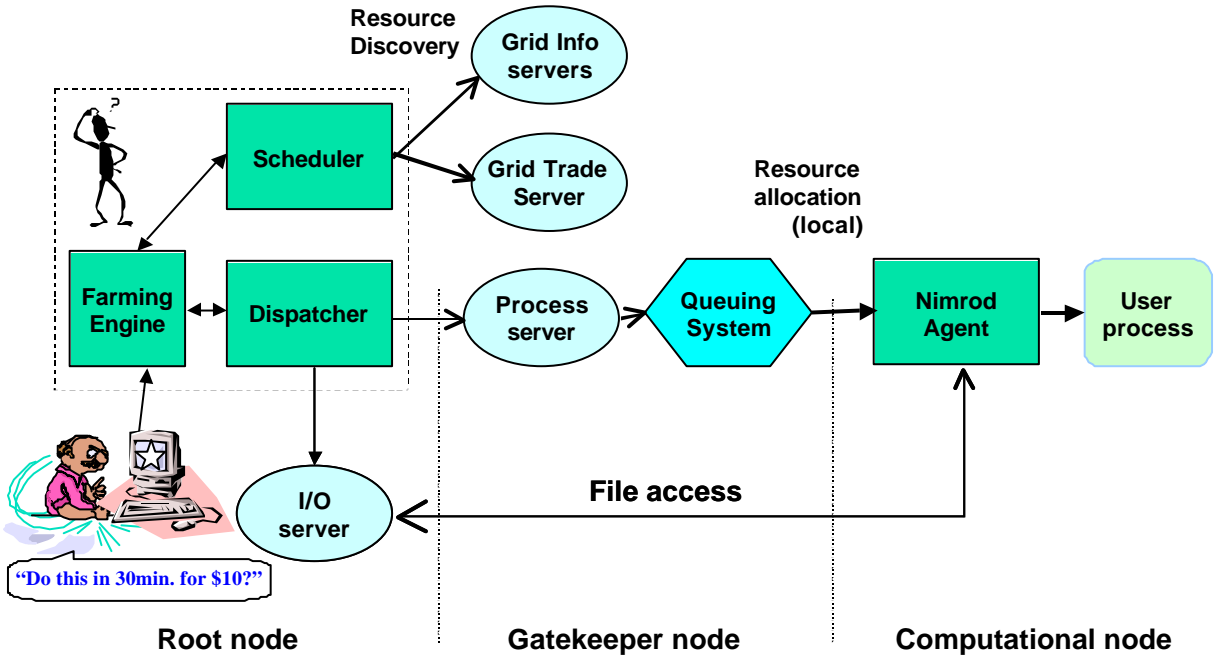

**Figure 4:** The Flow actions in the Nimrod-G runtime environment.

*The Task Farming Engine (TFE)*

The Nimrod-G task-farming engine is a persistent and programmable job control agent that manages and controls an experiment. It consists of a database to provide persistence which is accessed through a thin management interface. The farming engine is responsible for the parameterization of the experiment, the actual creation of jobs, the maintenance of job status, and provides the means for interaction between the clients, the schedule advisor, and the dispatcher. The TFE interacts with the scheduler and dispatcher in order to process jobs. It manages the experiment under the direction of schedule advisors, and then instructs the dispatcher to allocate an application task to the selected resource.

The TFE maintains the state of an entire experiment and ensures that it is recorded in persistent storage. This allows the experiment to be restarted if root node fails. The TFE exposes interfaces for job, resource, and task management along with the job-to-resource mapping APIs. Accordingly, scheduling policy developers can use these interfaces to implement other schedulers without concern for the complexity of low-level remote execution mechanisms.

### *The Scheduler*

The scheduler is responsible for resource discovery, resource trading, resource selection, and job assignment. The resource discovery algorithm interacts with an information service (the MDS in Globus), identifies the list of authorized and available machines, trades for resource access cost, and keeps track of resource status information. The resource selection algorithm is responsible for selecting those resources that meet the deadline and budget constraints along with optimization requirements. Nimrod-G provides three different scheduling algorithms [6].

### *The Dispatcher and Actuators*

The dispatcher triggers appropriate actuators to deploy agents on grid resources and assign one of the resource-mapped jobs for execution. Even though the schedule advisor creates a schedule for the entire duration based on user requirements, the dispatcher deploys jobs on resources periodically depending on load and number of CPUS that are available. We have implemented different dispatchers and actuators for each different middleware service. For example, a Globus-specific dispatcher is required for Globus resources, and a Legion-specific component for Legion resources.

### *Agents*

Nimrod-G agents are deployed on Grid resources dynamically at runtime depending on the scheduler's instructions. The agent is responsible for setting up the execution environment on a given resource for a job. It is responsible for transporting the code and data to the machine; starting the execution of the task on the assigned resource and sending results back to the TFE. Since the agent operates on the "far side" of the middleware resource management components, it provides error-detection for the user's task, sending the information back to the TFE.

The Nimrod-G agent also records the amount of resource consumed during job execution, such as the CPU time and wall clock time. The online measurement of the amount of resource consumed by the job during its execution helps the scheduler evaluate resource performance and change the schedule accordingly. Typically, there is only one type of agent for all mechanisms, irrespective of whether they are fork or queue nodes. However, different agents are required for different middleware systems.

## *2.4. Implementation and Technologies*

The Nimrod-G resource broker follows a modular, extensible, and layered architecture with an "hourglass" principle as applied in the Internet Protocol suite [11]. This architecture enables separation of different grid middleware systems *mechanisms* for accessing remote resources from the end user applications. The broker provides uniform access to diverse implementations of low-level Grid services. The key components of Nimrod-G, the Task Farming Engine, the scheduler, and the dispatcher are loosely coupled. The interaction among them takes place through network protocols. Apart from the Dispatchers and the Grid Explorer, the Nimrod-G components are mechanism-independent. The modular and extensible architecture of Nimrod-G facilitates a rapid implementation of Nimrod-G support for upcoming peer-to-peer computing infrastructures such as Jxta [25] and Web services [26]. To achieve this, it is only necessary to implement two new components, a dispatcher and an enhanced Grid Explorer. The current implementation of Nimrod-G broker uses low-level Grid services provided by Globus [9] and Legion [10] systems. We also support Nimrod-G dispatcher implementation for Condor [23] resource management system. The role of various Grid and commodity technologies in implementing Nimrod-G functionality and components is presented in Table 2.

While submitting applications to the broker, user requirements such as deadline and budget constraints need to be set and start application execution. These constraints can be changed at any time during execution. The complete details on application parameterization and jobs management information, starting from submission to completion, is maintained in the database. In the past the database was implemented as a file-based hierarchical database. In the latest version of Nimrod-G, the TFE database is implemented using a standard "relational" database management system.

The commodity technologies and software tools used in the Nimrod-G implementation include: the C and Python programming languages, the Perl scripting language, SQL and Embedded C for database management. The PostgreSQL database system is used for the management of the TFE database and its interaction with other components.

| Nimrod-G Functionality | Grid Services Used or Performed By |
|---|---|
| Application Model | Coarse Grained Task Farming, Master Worker, and Data Parallelism. |
| Application Composition | We support mechanism for application parameterization through parameterization of input files and command-line inputs for coarse-grained data parallelism. It basically supports coarse-grain, data parallel, task farming application model, which can be expressed using our declarative programming language or GUI tools. |
| Application Interface | The Nimrod-G broker supports protocols and interfaces that help in job management. Nimrod-G clients or problem solving environments can add, remove, and enquire about job status. They can set user requirements such as deadline and budget; start and stop application execution both at job level and the entire application level. |
| Scheduling Interface | The Nimrod-G broker supports protocols and interfaces that help in developing schedulers. The schedulers can interact with the TFE, inquire jobs, user constraints, and develop a schedule that maps jobs to resources. |
| Security | Secure access to resources and computations (identification, authentication, computational delegation) is provided by low level middleware systems like Globus. |
| Resource Discovery | Resource discovery involves discovering appropriate resources and their properties that match with users requirements. We maintain resource listings for Globus, Legion, and Condor and their static and dynamic properties are discovered using grid information services. For example, in case of Globus resources, we query Globus LDAP-based GRIS server for resource information. |
| Resource Trading and Market Models | The market-driven resource trading is performed using GRACE trading services. The Nimrod-G broker architecture is generic enough to support various economic models for price negotiation and using the same in developing application schedules. |
| Performance Prediction | The Nimrod-G scheduler performs the user-level resource capability measurement and load profiling by measuring and establishing the job consumption rate. |
| Scheduling Algorithms | Deadline and budget-based constrained (DBC) scheduling performed by Nimrod-G Schedule Advisor. Along with DBC scheduling, we support further optimization of time, cost, or surplus driven divide and conquer in scheduling. |
| Remote Job Submission | The Nimrod-G dispatcher performs deployment of Nimrod-G agents using Globus GRAM, Legion, or Condor commands. The agents are responsible for managing all aspects of job execution. |
| Staging Programs and Data on Remote Resources | In case of Legion and Condor it is handled by their I/O management systems. On Globus resources, we use http protocols for fetching required files. |
| Accounting (Broker Level) | Nimrod-G agents perform accounting tasks such as measuring resource consumptions and the scheduler performs the entire application level accounting. |
| Monitoring and Steering Tools | Nimrod-G Monitoring and Steering Client |
| Problem Solving Environments | ActiveSheets and Nimrod-O are enabled use Nimrod-G services. |
| Execution Testbed | The World Wide Grid (WWG) having resources distributed across five continents. |

**Table 2:** The Resource Broker Functionality and Grid Services Role.

## 3. Scheduling Experiments on the World Wide Grid Testbed

We have performed deadline and budget constrained scheduling experiments at two different times (Australian peak and off-peak hours) on resources distributed in two major time zones [7] using a "cost-optimization scheduling algorithm" [6] on the World Wide Grid (WWG) [15] testbed shown in Figure 5. Currently, the testbed has heterogeneous computational resources owned by different organizations distributed across five continents: Asia, Australia, Europe, North America, and South America. This testbed contains heterogeneous resources such as PCs, workstations, SMPs, clusters, and vector supercomputers running operating systems such as Linux, Sun Solaris, IBM AIX, SGI IRIX and True64. Further, the systems use a variety of job management systems such as Condor, RMS, PBS, and LSF. All these resources are grid enabled using Globus services.

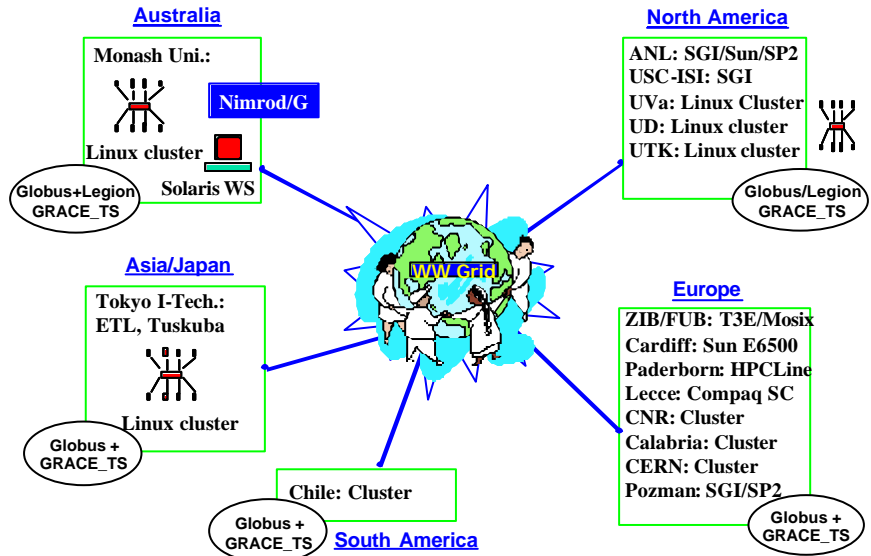

**Figure 5:** The World Wide Grid (WWG) Testbed.

We have performed an experiment of 165 CPU-intensive jobs, each running approximately 5 minutes duration. For a given deadline of 2 hours (120 minute) and budget of 396000 (G$ or tokens), we conducted experiments for two different optimization strategies:

1. Optimize for Time - this strategy produces results as early as possible, but before a deadline and within a budget limit.
2. Optimize for Cost -this strategy produces results by deadline, but reduces cost within a budget limit.

In these scheduling experiments, the Nimrod-G resource broker employed the *Commodity Market* model for establishing a service access price [8][]. It used grid resource trading services for establishing connection with the Grid Trader running on resource providers' machines and obtained service prices accordingly. The broker architecture is generic enough to use any of the protocols discussed in [8] for negotiating access to resources and choosing appropriate ones. The access price varies from one consumer to another and from time to time, as defined by the resource owners. Depending on the deadline and the specified budget, the broker develops a plan for assigning jobs to resources. While doing so it does dynamic load profiling to learn the ability of resources for executing jobs. Thus, it adapts itself to the changing resource conditions including failure of resources or jobs on the resource. The heuristics-based scheduling algorithms employed by Nimrod-G broker are presented in our early work [6].

We have used a subset of resources of the WWG testbed [15] in these scheduling experimentations. Table 3 shows resources details such as architecture, location, and access price along with type of Grid middleware systems used in making them Grid enabled. These are shared resources and hence they were not fully available to us. The access price indicated in the table is being established dynamically using GRACE resource trading protocols (commodity market model), but is based on an arbitrary assignment by us for demonstration purposes only.

| Resource Type & Size (No. of nodes) | Organization & Location | Grid Services and Fabric | Price (G$ per CPU sec.) | Jobs Executed on Resources | |
|---|---|---|---|---|---|
| | | | | Time_Opt | Cost_Opt |
| Linux cluster (60 nodes) | Monash, Australia | Globus, GTS, Condor | 2 | 64 | 153 |
| Solaris (Ultra-2) | Tokyo Institute of Technology, Japan. | Globus, GTS, Fork | 3 | 9 | 1 |
| Linux PC (Prosecco) | CNUCE, Pisa, Italy | Globus, GTS, Fork | 3 | 7 | 1 |
| Linux PC (Barbera) | CNUCE, Pisa, Italy | Globus, GTS, Fork | 4 | 6 | 1 |
| Sun (8 nodes) | ANL, Chicago, USA | Globus, GTS, Fork | 7 | 42 | 4 |
| SGI (10 nodes) | ISI, Los Angeles, USA | Globus, GTS, Fork | 8 | 37 | 5 |
| | | Total Experiment Cost (G$) | | 237000 | 115200 |
| | | Time to Complete Experiment (Min.) | | 70 | 119 |

**Table 3**: The WWG testbed resources used in scheduling experiments, job execution and costing.

The number of jobs in execution on resources (Y-axis) at different times (X-axis) during the experimentation is shown in Figures 6 and 7 for the time and cost optimization strategies respectively. In the first (time minimization) experiment, the broker selected resources in such a way that the whole application execution is completed at the earliest time for a given budget. In this experiment, it completed execution of all jobs within *70 minutes* and spent *237000 G$*. In the second experiment (cost minimization), the broker selected cheap resources as much as possible to minimize the execution cost whilst still trying to meet the deadline (completed in *119 minutes*) and spent *115200 G$*. After the initial *calibration phase*, the jobs were distributed to the cheapest machines for the remainder of the experiment. The cost of the time optimization experiment is much larger than cost optimization one due to the use of expensive resources to complete the experiment early. The results show that our Grid brokering system can take advantage of economic models and user input parameters to meet their requirements.

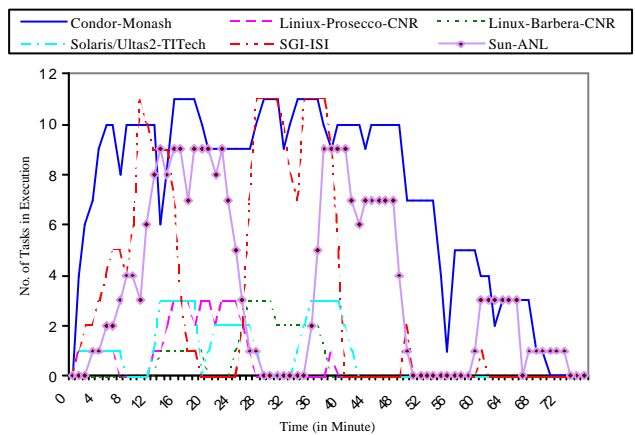

**Figure 6:** Resource Selection in Deadline and Budget based Scheduling for Time Optimization strategy.

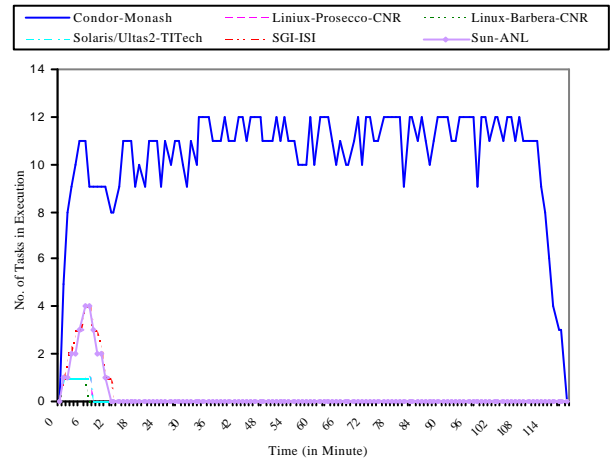

**Figure 7:** Resource Selection in Deadline and Budget based Scheduling for Cost Optimization strategy.

## 4. Conclusions and Future Work

The emerging grid computing technologies are enabling the creation of virtual organizations and enterprises for sharing distributed resources for solving large-scale problems in science, engineering, and commerce. The resource management and scheduling systems in grid environments need to be adaptive to handle dynamic changes in availability of resources and user requirements. At the same time, they need to provide scalable, controllable, measurable, and easily enforceable policies for management of resources. To address these requirements, we have proposed an economic paradigm for resource management and scheduling; and developed a prototype resource broker called Nimrod-G for scheduling parameter sweep applications on distributed resources.

The Nimrod tools for modeling parametric experiments are quite mature and in production use for cluster computing. A prototype version of Grid enabled tools and Nimrod-G resource broker have been implemented and they are available for download from our project web page. The Nimrod-G task farming engine (TFE) services have been used in developing customized clients and applications. An associated dispatcher is capable of deploying computations (jobs) on grid resources enabled by Globus, Legion, and Condor. The TFE jobs management protocols and services can be used for developing new scheduling policies. We have built a number of market-driven deadline and budget constrained scheduling algorithms, namely, time and cost optimizations with and without deadline and budget constraints. The results of scheduling experiments with different QoS requirements show promising insights into the effectiveness of economics paradigm for management of resources and their usefulness in application scheduling with optimizations. We believe that the computational economy approach for grid computing provides one of the essential ingredients for pushing grid into the mainstream computing.

In the near future, we plan to support scheduling with advance resource reservation. The economic models that will be used are driven demand-and-supply, tenders/contract-net, and auctions protocols [8]. These new models will require suitable scheduling algorithms to be developed. All of the models and associated scheduling algorithms will be evaluated through simulations and experimentations. In order to completely automate leasing of resources at runtime and online payments, we need digital currency, which is out of the scope of our work. However, we note that electronic currency technology is rapidly progressing with emerging e-commerce infrastructure [7]; we will incorporate them in Nimrod-G when they are available.